# To Explain Or Not To Explain: An Empirical Investigation Of Ai-Based Recommendations On Social Media Platforms

AKM Bahalul Haque, LUT University, Finland, bahalul.haque@lut.fi.

A.K.M. Najmul Islam, LUT University, Finland, najmul.islam@lut.fi.

Patrick Mikalef, Norwegian University of Science and Technology, Norway, patrick.mikalef@ntnu.no.

**Abstract**

AI-based social media recommendations have great potential to improve user experience. However, often these recommendations do not match the user interest and create an unpleasant experience. Moreover, the recommendation system being a black box creates comprehensibility and interpretability issues. This paper investigates social media recommendations from an end-user perspective. For the investigation purpose, we conducted a qualitative study with the users of the popular social media platform Facebook. We asked participants about social media content suggestions, comprehensibility, and explainability. Our analysis shows, users mostly require explanation whenever they encounter unfamiliar content and want to be informed about their data privacy and security. Furthermore, the users require concise, non-technical explanations along with the facility of controlled information flow. In addition, we observed that explanations impact the user's perception of transparency, trust, and understandability. Finally, we have outlined some design implications and presented a synthesized framework based on our data analysis.

## 1    Introduction

Social media platforms are increasingly adopting AI-based recommendation systems for personalizing content suggestions. The users of these platforms are also increasing rapidly and contributing to the platforms' metadata by sharing their created and modified content publicly, with friends, or within specific online communities (Guy et al., 2010; Wang et al., 2013). Moreover, the users can share their opinion about any topic, create and share their own portfolios, relationship status, and memories, and participate in various virtual communities worldwide. In a nutshell, these platforms have become a place for all types of virtual connectivity.

Social media platforms keep expanding as more and more content is being added to those platforms (Wang et al., 2013). Therefore, users often find selecting which content to interact with is challenging. The AI-based recommendation system has become a helper to solve this challenge and it is also one of the most important services in any social media platform. This service enables users to get in touch with suitable digital content such as news, vlogs, images, advertisements, etc. Personalized recommendations have recently started gaining popularity since they can provide users with tailor-made content such as news, articles, videos, images, and advertisements. This type of recommender system identifies individuals' characteristics, predicts their behavior, and delivers the required content. Moreover, the large amount of metadata helps to increase the efficacy of the personalized recommender system and, therefore, can lure in as many users as possible (Wang et al., 2013; Lu et al., 2015).

However, Due to the lack of explainability, the recommendations provided to the users raise concerns regarding the opacity of the whole system (Peake & Wang, 2018; Zhang & Chen, 2020). Explainable AI-based system is one of the potentially under-explored areas of social media recommendation systems (Haque e al., 2023; Laato et al., 2022). Most of the social media recommendation system-related studies to date focus on efficacy and user satisfaction (Guy et al., 2010), estimation accuracy (Bilgic & Mooney, 2005), the effectiveness of different types of explanations in context-aware applications (Lim et al., 2009), the intelligibility of instant messenger notifications and context-aware applications (Lim & Dey, 2011), the effect of accuracy on trust in machine learning model (Yin et al., 2019), explainable song recommendation (Zhao et al., 2019), explainable AI (XAI) based fake news (Chien et al., 2022) and hate speech detection (Mehta & Passi, 2022). The accuracy (performance) of the recommendation system is one of the performance metrics of model interpretability/explainability. The model's stated accuracy can be different in real-world use cases due to differences in training and real-world data. Therefore, the performance of the recommender systems algorithm might differ. Model accuracy has a significant impact on users' trust, however, that too also depends on users' domain knowledge and observation (Yin et al., 2019). Another performance metric is diversity, which indicates the diverse list of items a recommender system uses to generate recommendations. In this case, also there is a trade-off between diversity and accuracy in some algorithms such as in collaborative filtering (Bag et al., 2019).

Therefore, explainable recommendations in social media from an end-user perspective is a relatively underexplored area that can be an exciting study and contribute to explainable AI research (Haque e al., 2023; Laato et al., 2022). In this study, we have focused on AI-based content suggestions on Facebook to explore the explainability of the recommendation system. Facebook is one of the most popular social media platforms across different regions, having 2.9 billion monthly active users[1]. This platform enables users to share digital content, advertise products, establish community connections, etc., making it a suitable area for our investigation. However, along with the personalized recommendations in social media, especially on Facebook, the explanations are not readily available and easily accessible to its users. Moreover, though domain experts and machine learning researchers can get an initial idea about recommendations are made, it is very difficult for the end users (Zhang & Chen, 2020; Haque e al., 2023; Laato et al., 2022; Bilgic & Mooney, 2005).

Additionally, the need for explanation is dynamically on the rise due to various privacy policies and guidelines like GDPR[2] and guidelines from High level Expert Groups[3]. These concerns advocate the need to investigate the end user perspective of an explainable recommendation system. Therefore, our research objective is to address the following research questions:

RQ1. Why do users require explanations for AI-based content suggestions/recommendations in social media?

RQ2. How do the users want the explanations to be presented/ delivered to them and how it will impact their usage?

We have conducted semi-structured interviews among 30 regular Facebook users to investigate these questions. We asked the participants questions about their usage patterns, their opinion regarding the need for explanations with content recommended to them, and how it can impact their social media use. Analysis of the interview data shows that answer to RQ1 mainly revolves around (i) the content recommendation opacity and (ii) lack of clarity regarding users' personal data usage. Content recommendation opacity illustrates the fact that the users want to be informed about how the recommendations are made. The users often are startled if they come across unfamiliar recommendations. Moreover, the users believe that receiving an explanation along with the recommendation will help them understand how their personal data is being used and processed in the system. Therefore, explainable recommendations can potentially help reduce users' fear of online insecurity in terms of personal data usage. The answer to the first part of RQ2, "How do the users want the explanations to be presented/ delivered to them" reveals that the users require concise and non-technical information as explanations. In addition, they want to have a certain amount of control over the recommendation and the explanation because the users think this control will help the recommender system rectify their recommendation if necessary. Finally, analyzing the later part of RQ2, "how it will impact their usage" reveals that most users are yet to experience such explanations. However, according to their initial understanding, the users think explainable recommendations will positively influence their social media experience and usage.

---

[1] Global Social Media statistics, 2022. Available: https:/datareportal.com/social-media-users. Accessed: November 1, 2022

[2] General Data Protection Regulation. Available: https://gdpr.eu/. Accessed: 7th November 2022.

[3] Ethics Guidelines for Trustworthy AI. Available: https://ec.europa.eu/futurium/en/ai-alliance-consultation.1.html

The rest of the paper is organized as section 2 consists of the background that discusses a brief summary of social media recommendation, the basics of an explainable recommendation system, and related works; section 3 describes the research strategy which discusses the research methodology that is being followed, study area and design, data analysis, and reporting process, validity and reliability checking process; section 4 discusses the study findings which consists of demographic information, thematic analysis and mapping them with the research questions; section 5 discusses the implications which consist of design recommendations and synthesized framework for explainable recommendation interface design; section 6 discuss the limitation and section 7 concludes the paper.

## 2    Background

### 2.1    Social Media Recommendations

A recommendation system produces a certain set of suggestions using AI-based algorithms. Generally, a recommendation system can be personalized and non-personalized (Das et el., 2017). Personalized recommendation systems are mostly used in social media platforms. This type of system considers the user's previous behavior and interaction on social media (Seo et al., 2017). In the early days of social media recommendations, traditional recommender systems were used to provide recommendations using content-based filtering, collaborative filtering, etc. (Shapira et al., 2013; Wang et al., 2013). However, social media users and the amount of data have increased in the last decade. Moreover, more functionalities are being added to social media to increase versatility. The users are provided with various recommendations, such as friend suggestions, group recommendations, product recommendations, Facebook page recommendations, etc. (Caers et al., 2013; Baatarjav et al., 2013). These different types of recommendations are provided by social media using machine learning algorithms such as deep learning, which is a layered combination of many large and complex neural networks (Guy et al., 2010; Geng et al., 2015). These algorithms also provide recommendations based on the user's interaction with various features of social media and social connectivity. Therefore, users' social connection and activity are being fed into the complex machine learning models, to provide the users with the best possible recommendation. Since these models are inherently opaque even to experts, it is more difficult for lay users to comprehend how a specific recommendation is being made (Peake & Wang, 2018; Zhang & Chen, 2020; Haque et al., 2023).

Explainable recommendation facilitates the target user to understand why and how a recommender system recommends any particular content (Zhang & Chen, 2020). These recommendations promote transparency, persuasiveness, efficacy, trustworthiness, and satisfaction. Many explainable recommendation approaches, notably model-based methods, have been suggested and deployed in real-world systems in recent years. Explanations can be delivered in two ways; post-hoc (model agnostic) and directly from an explainable model (model intrinsic). Model intrinsic explanations can also be termed interpretable models, which can deliver explanations for the prediction along with the decision-making. It means the decision-making mechanism of the model is naturally transparent since the models are inherently interpretable. Post-hoc (model agnostic) explanations are not inherently explainable; instead, the model remains as black box, and the explanations are produced after the decision is rendered. According to Lipton (2018) and Miller (2019), the explainable recommendation types mentioned above refers to the understanding of the topic of cognitive psychology (Lipton, 2018; Miller, 2019). Model intrinsic explanations can relate to the fact that, while making decisions, we as humans carefully consider all the traditions and find appropriate logic to know why we made this decision. Model agnostic explanations can relate to the fact that we try to produce explanations after we have decided on something. This technique helps to justify or reason against the decision that we have made (Lipton, 2018; Miller, 2019). The concept of a personalized recommendation system started long back, the "explainable recommendation" terminology was first used by Zhang et al. (2014). However, explainable AI has gained much attention in recent years, emphasizing the importance of explainable recommendation systems (Gunning, 2017). The explainable AI umbrella covers a wide range of terminologies and concepts of AI, such as recommender systems, autonomous systems, computer visions, data-driven decision-making, reinforcement learning, interpretable machine learning, natural language processing, etc. Therefore, an explainable recommendation system is closely related to explainable AI.

## 2.2    Related Works

Explainability content recommendation systems have been analyzed from various perspectives, such as content-based filtering and collaborative filtering techniques. These two are the most popular types of recommendations system (Bilgic & Mooney, 2005). In addition, researchers have investigated the human comprehensibility of these systems in terms of improving the user experience (Lim & Dey, 2009). Explainable recommendations have also been studied from context-aware intelligent systems context (Lim & Dey, 2009). The researchers investigated the intelligibility of various context-aware systems and how the user interacts with them. Similarly, during the same time, Lim et al. (2009) investigated how counterfactual explanations impact the understanding of recommendations from an end-user perspective (Lim et al. 2009). In addition, explainable recommendation system algorithms can reduce the effect of unfairness and, at the same time, produce correct recommendations (Fu et al., 2020).

A personalized explainable recommendation system attracts more users. Such as Zhao et al. (2019) designed an explainable song recommendation system that brought in more end-user and increased understandability (Zhao et al., 2019). He et al. (2015) proposed collaborative filtering and view-based explainable AI techniques that generated more transparency among the users. In that system, they used user reviews and ratings to identify the aspects of a specific item and used a ranking algorithm for personalized recommendations (He et al., 2015). Cheng et al. (2019) used a similar technique but added the item images to enhance the product recommendation (Cheng et al., 2019). Wang et al. (2018) proposed a post-hoc explainable recommendation system based on reinforcement learning. The model can also explain collaborative filtering recommendations and has proved to be more beneficial than baselines

(Wang et al., 2018). Xian et al. (2019) have also used reinforcement learning, but their approach has additionally used a policy-guided graph search strategy and soft rewards strategy (Xian et al., 2019). Collaborative filtering is one of the most widely used techniques in a personalized recommendation; since it does not provide enough explainability, Wang et al. (2018) used embedding-based methods and an easy-to-interpret attention network so that suggestions become explainable (Wang et al., 2018). Tan et al. (2021) have used the counterfactual explanation technique to design an explainable recommendation system. In their proposed system, two perspectives have been used (Tan et al., 2021). The first one regards the user's interest in the suggested item, and the second is why the algorithm suggests the product.

The discussion of related work shows that the previous works mostly followed a more technical approach in explaining recommendations. The researchers have focused on the model interpretability, model agnostic behavior of recommender systems, explainability of reinforcement learning models, etc. These works primarily focus on various recommender systems' intrinsic structure, design, and efficiency. Moreover, research has been conducted from end users' points of view regarding user understandability, transparency, usability, etc. However, very few researches have been conducted in the social media context and explicitly in what context explaining the recommendation becomes vital and how the user reacts to it.

## 3    Research Strategy

### 3.1    Study Area and Design

Our research aimed at explaining the social media recommendations that also fall under explainable AI's umbrella. We have focused on Facebook since it is the most common and widely used social media platform. Furthermore, the recent controversies related to Cambridge Analytics' data privacy, and news controversies (Mitra & Khosrowshahi, 2021; Crocco et al., 2020; Carlson, 2018), this social media platform present a situational context to investigate the end user's perspective of content suggestions. Facebook allows people to connect globally using messages, voice, and video calling features and sharing different content. Various Facebook groups and pages are built to connect communities and people with similar and special interests. It has also become a popular spot for a marketplace where advertisers can promote their products. Due to the veracity of content on this platform, users get recommendations for various items. This study investigates these content suggestions and their explainability from the lay users' perspective.

In this study, we have asked the participants semi-structured interview questions. The participants are regular Facebook users. We asked them questions regarding their usage patterns, their view and need for explainable recommendations, and how it can affect their usage. The rationale behind asking these questions is related to the nature of Facebook content suggestions. For example, from our regular use of Facebook, we can see that content suggestions vary from person to person. Sometimes, the suggestions are relevant to our day-to-day usage. However, in some cases, the suggestions are totally out of the ordinary and unexpected. To make the users understand and get a clear picture of content suggestions, we also used several examples such as content suggestions in various e-commerce websites, movie streaming websites, travel destination recommendations, etc. Often these websites provide users with product recommendations and some type of hints such as "frequently watched movies"/ "previously purchased products" etc. These examples helped the users get an overview of what we meant by content recommendation, especially the explainable recommendations. So, we first asked the participants about their Facebook usage patterns and purpose. Later we wanted to know about the content they regularly come across while using the system. These questions provided useful insights regarding the recommendation differences and similarities among users. Then we asked the participants about their thoughts on explainable recommendations and how they want them. These questions helped users understand the context and discussed the need for explanations, explanation format, and how much information should be in the explanation. Later, we asked the participants to imagine that they were presented with their expected explanations and, in that case, how it would impact them. We wanted to get an initial response from them because long-term exposure to explanations is not within the scope of this study.

### 3.2     Data Collection and Analysis

For the data collection, we conducted 30 in-depth semi-structured interviews with regular Facebook users who spend at least two hours daily. In this study, we have used semi-structured interviews because of their potential to shed light on pertinent, real-world questions related to the studies' theoretical underpinnings (Maxwell, 2008; Catterall, 2000). In addition, this approach facilitates the researcher to acquire more focused and specific information regarding a specific topic (Hillebrand & Berg, 2000). At first, we designed and developed the questionnaire based on two sample interviews. From the sample, we found that the term "explainability" of explainable recommendations was not a common term to the users. Therefore, we used the term "reasoning,"/ "hints,"/" explanations" during the interview. As a direct outcome of the sample interview, we constructed and revised the questionnaire and adjusted our overall interview criteria. We ensured all of our respondents at the outset that their privacy would be safeguarded and their responses would remain anonymous so they would feel comfortable answering every question. Therefore, we abstained from requesting any type of personal identification information. In most cases, the participants were not interested in identifying themselves with personal information in the recording. Therefore, we have recorded the interview responses related to this research only.

The participants were invited through social media groups and pages. We managed to get 30 participants who volunteered to participate in the interview. The interview sessions were conducted both offline and online using zoom. We have taken notes during the session. The data analysis followed a thematic map; therefore, we identified the codes based on the recorded transcripts. The second step consisted of developing patterns by systematically combining codes with identical components. We found three themes and several subthemes and aligned them with the research questions of our study.

### 3.3    Validity and Reliability

The qualitative interview data were analyzed through a thematic analysis using the inductive technique (Braun & Clarke, 2006). Thematic analysis is a technique for extensively and methodically analyzing qualitative data. The information was then compiled and triangulated in accordance with data type and attributes. Triangulation procedures were applied by researchers to confirm the reliability, validity, and dependability of their data (Golafshani, 2003). In this research, investigator triangulation has been applied where multiple investigators were involved in conducting and overseeing the whole procedure. (Denzin, 1978). The primary author oversaw the entire data analysis procedure, beginning with the creation of interview questions, data collection, and analysis. Later the other authors reviewed and checked the data collection and analysis procedure. Finally, all the authors agreed on this study's data analysis and findings. The findings are then discussed and reported in the manuscript. Moreover, a series of pre-test questionnaires were prepared to evaluate the questionnaires' reliability. In this study, participants volunteered for their interviews. We have maintained their personal data privacy and anonymity throughout the whole research process.

## 4      Study Findings

This section outlines the findings of the interviews through data analysis. The findings represent various aspects of end-user-oriented explanation of social media content recommendation. We have categorized the findings according to the research questions in this study's introduction section.

However, the demographic information of the participants is mentioned as follows. According to Table 1, more than half of the participants are male, and the rest are female. Most participants fall between the ages of 18-24, and a considerable number fall between the ages of 25-35. The participants were mostly students; among them, 10 were undergraduate students, 6 were graduates, and 3 were postgraduate students/doctoral students. Two participants were involved with the industry jobs.

| Gender of the Interviewees | Male: 21 |
| | Female: 9 |
| Age Group | 18-24: 16 |
| | 25-30: 11 |
| | 31-40: 3 |

Table 1. Demographic Information of the Participants

### 4.1      Thematic Analysis Aligned with Research Question

This section outlines the thematic analysis of the interview data. We have also aligned these themes and subthemes with this study's research questions. A summarized overview is shown in Figure 1.

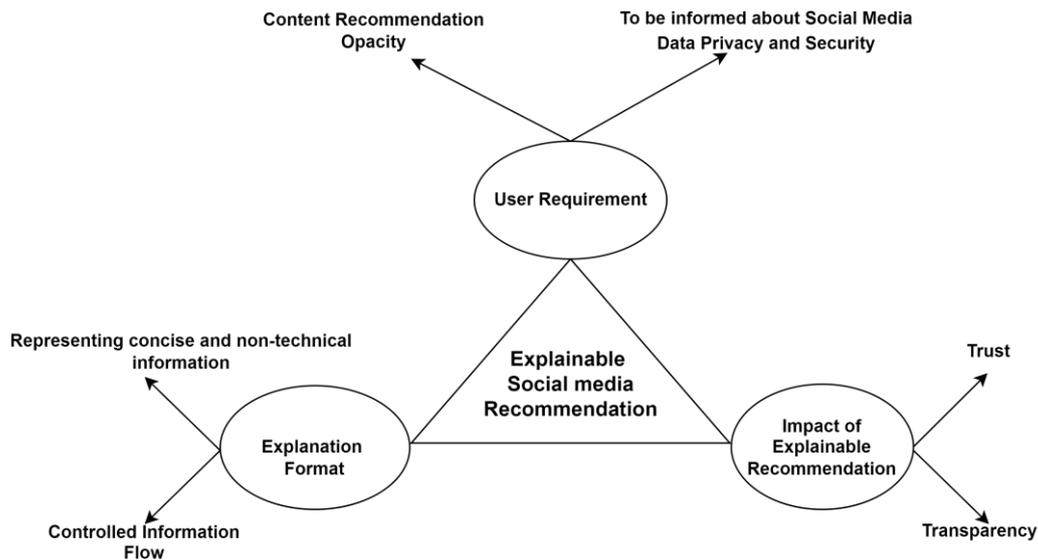

Figure 1. Thematic Analysis and Mapping with the Research Question

4.1.1    Addressing RQ1

**Content Recommendation Opacity**

Social media users have increased during the past few years. In the interview session, we tried to understand the nature of social media recommendations from the participants' responses. However, most users said they regularly come across various unfamiliar content. Generally, the content suggestions match their own interest, but not always. Therefore, the participants would like to know why they come across both these types of content. Most of the time, they do not have any clue why those contents appeared on their homepage. One of the participants said,

"I mostly use Facebook for connecting with facebook and interacting in different facebook groups. Often, i get suggestions of various facebook groups. However, there are some content which I did not even search in facebook but i can see them in my homepage. I want to know know they are coming to my facebook and on which basis."

The participants also mentioned seeing inappropriate content (videos, images, etc.) on their Facebook page, and it becomes really annoying and uncomfortable for them to see those while surfing the internet. The situation becomes more uncomfortable in front of friends and family. That's why they do not want to be under the blanket of ignorance. They feel Facebook should explain the reasons to them. One of the participants said,

> " I generally watch videos on Facebook to spend my leisure time. Sometimes I see some pictures or reel video and after that, Facebook shows me that type of thing; sometimes it provides some bad recommendation, for example, I saw a reel video that is slightly adult that have some slung or adult seen, facebook then automatically recommend that type of videos, that is very uncomfortable for me. So, I want Facebook to answer me why they are showing me these videos"

In addition to these responses, the users provided their opinion on recommendation transparency and providing control over their data. This information can be included in the explanations for making the system more transparent. According to one of the respondents,

> "I think Facebook should be more transparent from which sources they made the recommendation and should also give users more control over their data, what to share to the public what not."

Therefore, Facebook users require explanations so that they can make informed decision making. The participants talked about being "frightened" of social media due to its suspicious activity. Hence, they do not want to be in the dark regarding content recommendations.

**To be Informed about Social Media Data Privacy and Security**

Human minds naturally fear things they cannot see or understand. We observed from the interview that Facebook users are skeptical of their online security, especially regarding their personal data. Users often come across recommendations that they have not searched on Facebook but may be in any other browser or search engine. They start to get surprised whenever they see this type of thing on their social media, and it also intensifies their preconceived idea that their personal data is used by every social media platform and search engine. One of the respondents said,

> "I can understand the product recommendation that I repeatedly look for on Facebook. But, I do not understand how Facebook is able to show me what I am thinking in my mind or what I discussed with my friend. I am sure, I did not searched it in Facebook, may be I searched in google. Still, I see it in Facebook. It keeps on recommending me over and over again. How did it came to know."

Some respondents are afraid of their data being stolen. They think Facebook is stealing their data from their profile and usage logs. For this reason, they need to know what is happening behind the curtains of algorithms, and Facebook should provide users with appropriate explanations for that. One of the respondents said,

> " It will not make most people think a platform is "stealing" info from me for whatever unethical unexplained motives."

Furthermore, the participants also expressed their concerns about being tracked online. This fear is primarily fueled by online news portals and blog posts. These types of news have made them scared about their activity being tracked online. So, they think they are being tracked online whenever they use Facebook. One of the participants said,

> " What I believe is we are living in a world where our every actions/response have a virtual trace. This virtual presence is traded between giant tech companies with or without our consent. Even what we browse on the chrome browser pops up in our Facebook advertisement. The feeling of I am being traced makes me feel insecure."

However, explaining the recommendations can remove their dilemma. The users are not certain whether social networking sites are tracking them. However, including explanations along with recommendations, in fact, can shed some light on the real scenario. Therefore, it is understandable from the user's opinion that they hold many misconceptions and confusion regarding how social media collect and process their personal data. These misconceptions make users feel insecure about their online activities.

### 4.1.2 Addressing RQ2

In this section, we will address our second research question. While conducting the interview, we observed that most participants were concerned about getting explanations since they could not understand why certain recommendations were shown to them. However, their concern increases when the recommendations do not match their expectation or if they are not familiar with the recommendations. In this research objective, we will analyze how Facebook users want the recommendations to be explained.

**Representing concise and non-technical information**

Users are concerned about how social networking platforms, including Facebook, use their data to provide recommendations. They have a common idea that these websites utilize usage history to provide them with recommendations. One of the respondents said,

" I know that all this happens due to data sharing. Most of the apps used in mobile phones sell user data to Facebook. For Example, when you browse something in YouTube the next day you come across advertisement regarding that topic. It is because of the filtering algorithm of Facebook."

However, the recommendations are not always up to the mark for the end users. Often, they come across boring recommendations. One of the respondents said,

" Sometimes I'm feeling better because, this contents guidelines usually beneficial for the general population as a whole. But sometimes few advertisements is so disgusting while I'm feeling bored."

In the case of those boring and annoying recommendations, the "explanation need" from the user end increased. Participants said, as a user they would like to have a prompt explanation that will show why they repeatedly see these contents. Furthermore, they would also like a brief and concise explanation. As one of the respondents said,

" There must be a valid reason behind suggesting an ad; Otherwise, as a Facebook user, I find it annoying. Whenever an ad is shown, it should immediately explain why the ad is being shown."

The respondents also said they require explanations that include the content's aim and objective. In addition, a proper summary should include as part of the explanation.

" In my opinion, some contents outline, topic name, aim & purposes and finally executive summary should be provided along with those contents."

Other respondents also emphasized the importance of providing brief but understandable details in the explanation. One of them said,

" For me, a brief explanation would be sufficient. E.g. "This group is recommended to you because you are also part of group XYZ which is similar to this group and has 8 same members "

Including the concise explanation, users require non-technical information in the explanation. Users think that technical explanations may not be comprehensible by lay users.

**Controlled Information Flow**

As the data analysis shows, the users preferred customized and controlled information flow. Social media already contains a lot of information on the home pages. Amongst those explanations, if the explanation flow is not controlled, it might create problems for the users with understandability. One of the respondents said,

> ".... I do not mind them basing suggestions of these things as I'm getting curated content. However, there should be an option for them to not automatically access such things as other may not be as liberal in this context…."

Another opinion by the respondents was the explanations could be shown to them upon permission from the users. The users think it will provide more control on the user end. According to the user's opinion, the explanation of recommendations can be conveyed whenever the users want it, making the whole explainable recommendation system more controllable for the users. Here, we can observe that the users would like to have on-demand services from the explainable system since it will make them feel in control.

> "..the recommendation could be explained on permission, such as what would be recommended to me can ask a permission for the recommendation on/off, because sometimes it gets annoying always getting the recommendation. It could be controlled by the users…"

Moreover, the users also would like to have specific choices regarding the recommendations and explanations. For example, one of the respondents said

> " Facebook should give a choice whether the user wants to view the advertisement or not."

Controlled information flow for explanation will also facilitate the user's interaction with the important news and contents. If the users are provided with too much information, it may hamper user interaction as said by a user,

> " I think the kind of content we look for on Facebook and chat with friends; Facebook's artificial intelligence suggests content to me based on that. This annoys me because the same kind of ads keep popping up on the timeline. By doing this, the important posts fall behind; can't see"

These opinions of users, along with others, actually provide us an important hint that Facebook recommendations and their explanations should have more user control rather than providing default explanations and default view mode.

**Impact of Explainable AI-based Recommendation**

We asked the participants about their perceptions and if they were presented with their expected explanations. In response to these questions, some participants said. They do not know how to react since they have not yet interacted with such a system. Their opinion will depend on the explanation they receive, and they can answer the question when they encounter such a system in real life.

However, some participants said their usage patterns would change when they received explainable recommendations. For example, they can choose which content to interact with and which to ignore. Along with the explanation, they also think the system should provide more user control so that the ignored recommendations and explanations should not appear in the future. Holistically, an explainable system will impact their overall trust in the system. One of the participants said,

> " I will be able to find similar content with more ease if I understand how the recommendation works a little bit. I can also trust the platform more due to transparency."

> " Well, Facebook could really assist me out if it gave me good reasons for its suggestions. Simply said, I'll be able to access the material I care about. In order to ensure that we continue to receive the stuff that I enjoy, I will enforce some form of logical usage restrictions."

Social media users spend a lot of time on those platforms. They think unnecessary content suggestions and interaction kill their time. If they receive a proper explanation for the contents they come across, it will be helpful for them to choose the right one and manage their time. One of the participants said,

> "My usage become relevant because I roam around those group/page which can help me, benefit me. Otherwise I often kill my time watching unnecessary things on fb."

Explainable recommendations can have both positive and negative impacts. The users think there will be two types of users who will find this extremely useful as they can customize their social media experience. On the other hand, there will be other people who will find this system boring and find the explanations unnecessary. According to a user,

> "If a person is choosy or judgmental, I think it will have a positive impact. But it will have both sides of impact as well. So choosy people like me will find it positive, on the other hand, not-so-choosy people may find this feature boring."

We observed that the users tried to portray that providing an appropriate explanation will certainly change their perception; however, they just don't know how much since they have yet to interact with explainable social media content recommendation systems. Therefore, the participants reflected upon their immediate perception of such a system.

# 5     Discussion and Implication

In this paper, we have investigated the end users' perceptions of explainable recommendations in social media platforms. We have conducted semi-structured interviews to dig deeper into AI-based recommendations and collect qualitative data for analysis. Later, we analyzed the results into thematic categories and aligned them with the research questions. This section reflects upon those analyses and outlines the implications.

## 5.1    Collect User Preference on Runtime

The users are generally curious about social media recommendations and want to know how they appear on the homepage. However, the user's demand for an explanation is triggered instantly if they encounter any irrelevant recommendation. As the participants explained, they mostly come across content suggestions related to browsing and search history. However, users often come across irrelevant content suggestions and instantly look for appropriate reasons. The possible reason for irrelevant content suggestions can be the constant increase of metadata, algorithmic inefficiency, diverse functionalities, social media services, etc. However, the demand for explanation increases if the user is exposed to unknown and unfamiliar content on their social media. Another stimulating factor to explain AI-based recommendation is online security and privacy. Therefore, to feel protected and familiar with their personal data processing, the users require explanations for the social media recommendations, irrespective of familiar or unfamiliar ones.

Therefore, the design implication for AI-based recommender system designers can be collecting users' opinions on whether the user finds a specific recommendation familiar or unfamiliar. After collecting the responses, the system can provide choices to the users on whether they would like to view the explanation. This information is crucial since it will be difficult for the developers to design the system without knowing how explainable the interface should be. Bunt et al. 2012 discussed the dilemma of on-demand explanation about "why" users require explanation (Bunt et al., 2012). However, our investigation shows that users mostly require explanations whenever encountering unfamiliar content suggestion. Therefore, as a design recommendation, the developer should include the feature of giving users an option to select whether the recommendation is familiar or unfamiliar. This way, users will not have to engage in multiple interactions with the system but will get their required services automatically. Moreover, as part of the explanation of social media recommendation, information related to personal data collection and processing should be included explicitly so that the users feel safe about their online presence and interaction. Therefore, along with the algorithmic working principle (Branley-Bell et al., 2020; Ehsan et al., 2019) and contextual information (Daudt et al., 2021) as part of the explanation, personal data collection and processing information can be a vital part.

## 5.2     Duality of Explainable Recommendation

Explaining AI-based content suggestions on social media platforms can have positive and negative impacts. An explanation will help people understand the recommendations and increase understandability and transparency. However, on the other hand, the users have also talked about presenting the explanation in a concise or summarized format. Moreover, they think there should be a controlled flow of information as an explanation. Otherwise, too much information will annoy the user and create a cognitive overload. These factors(need as much information possible to understand the content suggestion and controlled information flow) are two opposite sides of the explainable recommendations which can also be termed the duality of explainable recommendations. However, the main hurdle for system developers and designers is to eradicate the duality. To do that, they need to understand how much information is "enough" for a user.

Interestingly this requirement can be different from user to user. As T. Miller (2019) explained, cognitive biases exist while a user selects and evaluates the explanation (Miller, 2019). Moreover, the information required as an explanation differs among users (Miller, 2019). Our investigation also observes a similar scenario regarding social media recommendations. Therefore, the design implication can be that the user interface should be able to dynamically allocate explanations according to the user's choice. The dynamic allocation of explanation can be based on the user choice on runtime. The system can learn from the user choice and design personalized recommendations in the future.

## 5.3     Synthesized Framework for Explainable AI-based Recommendation System

We synthesized a theoretical framework (see Figure 2) from the interview data analysis. This framework shows that the user's need for explanation stimulates how explanations should be provided. As the explanations are provided in the correct format, it will positively impact the system's trust, transparency, and usability and ultimately provide a better user experience. However, in this case, the correct explanation format is yet to be known for social media recommendations. Researchers can use this synthesized framework to conduct empirical works to identify how social media recommendations can be better explained to the users and how those explanations will affect user perception.

The synthesized framework proposed in this section is not a complete development cycle, however, practitioners can gain vital insights from it. Our observation from data analysis reveals that the users mostly need explanations if they come across unfamiliar content, feeling of online insecurity and their curiosity regarding the recommender system. These identified problems stimulate the need for collecting the design requirements. However, apart from the "identified problems" in this work, several others can also exist, if research is done from a broader perspective or considering other AI-based systems. Our analysis in this paper shows various design requirements such as explanations should be concise and summarized, non-technical descriptions and other attributes related to explanation formats. These design requirements can later be used to design and develop an explainable recommender system interface. After developing the system, it should be implemented in the platform for evaluation. Since social media platforms accommodate a wide variety of users, the evaluation process should take a reasonably long time. Moreover, the evaluation should happen in runtime, so that the social media service is not interrupted. Users can provide feedback as they use the system. This feedback can be incorporated into the system after the developer reviews it and prioritize it.

Furthermore, since the recommender system is AI-based, it will also learn over time from user interaction (Das et al., 2015; Glikson, & Woolley, 2020). This learning process will be throughout the product's lifetime, as long as the recommender system remains with the social media platforms. Therefore, to our understanding, most improvement iterations should happen automatically with the learning algorithms with time (Jüngling et al., 2020). Other things such as changes in various modules of the explainable interface, will need human intervention, and they will have to go through the iterative process of design, development, deployment, and evaluation. Finally, we will be able to get a system that can explain AI-based recommendations to the users. From our analysis, we observe that at the end of all the steps explainable recommendation system and interface should promote trust, better usability, transparency, and users time management will be optimized. AI-based system developers can use this framework and be test in various scenarios to evaluate and modify it. The evaluation can provide vital implications to the practitioners for future work. In our future work we also plan to evaluate frameworks synthesized from this study through from professional opinion and direction.

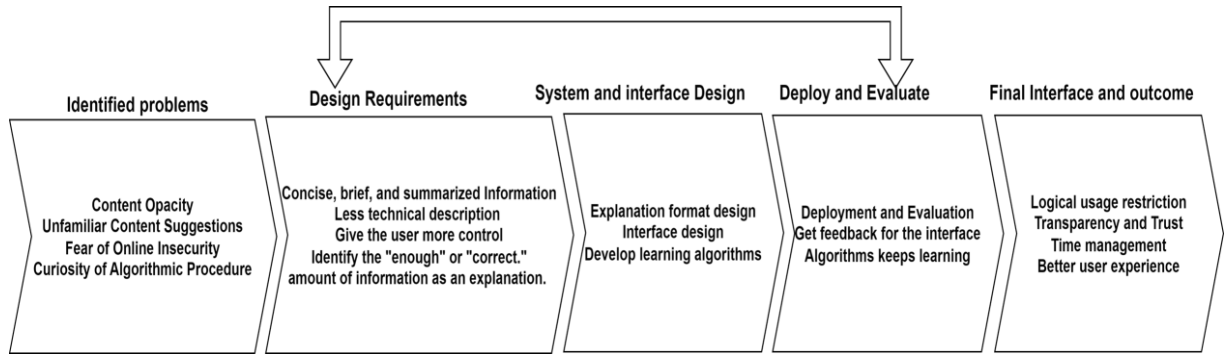

Figure 2: Synthesized Framework for Explainable AI-based Recommendation System Development

## 6       Limitations and Future Research

Our empirical research investigated explainable social media recommendations from a user perspective. The users provided valuable insights about how social media impacts their life, why they use it, and how explainable recommendations can help them. However, since explainable recommendations is a relatively new terminology for the end users, their comprehension level is still relatively low. Moreover, users are also, in some cases, unaware of any algorithmic explanation feature coming on social media. Therefore, longitudinal studies and presenting the users with a more visual representation of how recommendation works in social media can reveal diverse user feedback and produce valuable insights. In addition, in this study, the study participants are primarily young, aged between 18 to 40. In future studies we would like to add more participants who are in their 40s, 50s, 60s, and even more, if possible. Expanding the participants' age group can add new insights. In addition, we will extend our work by evaluating the proposed framework with AI professionals.

## 7       Conclusion

In this paper, we have investigated the topic of explainable AI-based recommendations in social media platforms from an end user's point of view. The topic can be brought under the umbrella of explainable AI. This work presents an in-depth qualitative analysis of social media recommendations. The analysis shows that users need explanations for content recommendations to be informed about the system and protect their online privacy and security. The investigation reveals the user's required explanation format and how explainability will affect the users. Though we have used the example of Facebook here, the findings can also be matched with other social media platforms. Our work highlights some crucial design implications for the developers and for the researchers as well. These design implications can be adopted for further experimental and empirical study to develop a robust framework for user-centric explainable AI systems.